\documentclass[prl,nofootinbib,superscriptaddress,twocolumn]{revtex4}
\def\mysection#1{{\bf #1.} }
\def\mysections#1{{\bf #1.} }

\usepackage{amssymb}
\usepackage{amsmath}
\usepackage{graphicx}
\usepackage{longtable}
\usepackage{verbatim}
\usepackage{amsfonts}
\usepackage[bookmarksdepth=2]{hyperref}

\newcommand{\be}{\begin{eqnarray}}
\newcommand{\ee}{\end{eqnarray}}
\newcommand{\bea}{\begin{eqnarray}}
\newcommand{\eea}{\end{eqnarray}}
\newcommand{\beq}{\begin{eqnarray}}
\newcommand{\eeq}{\end{eqnarray}}
\def\beqa{\begin{eqnarray}}
\def\eeqa{\end{eqnarray}}
\newcommand{\no}{\nonumber}
\def\lsim{\mathrel{\rlap{\lower4pt\hbox{\hskip1pt$\sim$}}
    \raise1pt\hbox{$<$}}}         
\def\gsim{\mathrel{\rlap{\lower4pt\hbox{\hskip1pt$\sim$}}
    \raise1pt\hbox{$>$}}}         

\begin{document}

\vspace*{-100mm}

\title{\boldmath AMS02 results support the secondary origin of cosmic ray positrons}

\author{Kfir Blum}\email{kblum@ias.edu}
\affiliation{Institute for Advanced Study, Princeton 08540, USA}
\author{Boaz Katz}\email{boazka@ias.edu}
\affiliation{Institute for Advanced Study, Princeton 08540, USA}
\affiliation{Bahcall Fellow}
\author{Eli Waxman}\email{eli.waxman@weizmann.ac.il}
\affiliation{Dept. of Part. Phys. \& Astrophys., Weizmann Institute of Science, POB 26, Rehovot, Israel}

\vspace*{1cm}

\begin{abstract}
\noindent
We show that the recent AMS02 positron fraction measurement is consistent with a secondary origin for positrons, and does not require additional primary sources such as pulsars or dark matter. The measured positron fraction at high energy saturates the previously predicted upper bound for secondary production \cite{Katz:2009yd}, obtained by neglecting radiative losses. This coincidence, which will be further tested by upcoming AMS02 data at higher energy, is a compelling indication for a secondary source. Within the secondary model the AMS02 data imply a cosmic ray propagation time in the Galaxy of $< 10^6$~yr and an average traversed interstellar matter density of $\sim 1~\rm cm^{-3}$, comparable to the density of the Milky Way gaseous disk, at a rigidity of $300~\rm GV$.
\end{abstract}

\maketitle

\noindent
\mysection{Introduction}
\noindent
The AMS02 experiment announced a new measurement of the positron fraction (ratio of $e^+$ to total $e^\pm$ flux) in Galactic cosmic rays (CRs)~\cite{PhysRevLett.110.141102}. The new measurement extends to high energy, $E\sim350$~GeV, with precision significantly superseding earlier experiments~\cite{Adriani:2008zr,Adriani:2010ib,FermiLAT:2011ab}. The positron fraction is found to increase with energy, apparently saturating at $e^+/e^\pm\sim0.15$ at $E\sim200$~GeV.

A rising positron fraction stands in conflict with expectations based on popular diffusion models, assuming a homogeneous diffusion coefficient and a cosmic ray halo scale height that is independent of cosmic ray rigidity (see, e.g.~\cite{1982ApJ...254..391P,Moskalenko:1997gh,Delahaye:2010ji}). This conflict has triggered numerous analyses invoking hypothetical primary sources for the positrons such as pulsars and annihilation or decay of dark matter particles.

\begin{figure}[!h]\begin{center}
\includegraphics[width=0.5\textwidth]{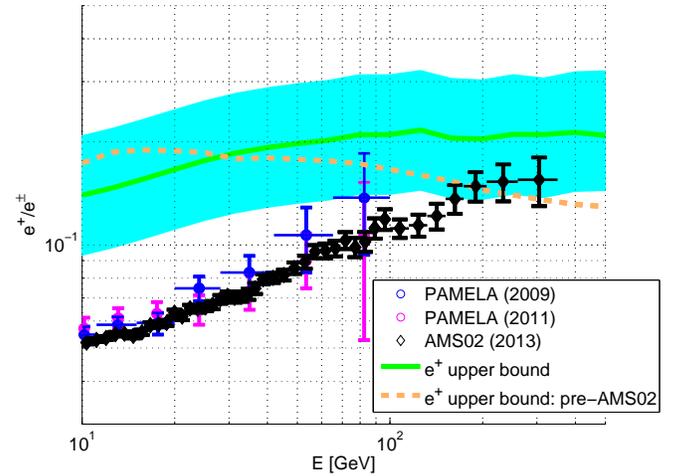}
\end{center}
\caption{
Positron flux upper bound vs. data, presented in terms of the positron fraction. The theoretical $e^+$ upper bound, divided by the $e^\pm$ flux measured by AMS02~\cite{AMS02ICRC}, is given by the green line. The cyan band shows the estimated calculation uncertainty. The calculation here is identical to that of Ref.~\cite{Katz:2009yd}, but uses the most recent B/C and $e^\pm$ data from AMS02~\cite{AMS02ICRC}. The result of the same calculation using pre-AMS02 data (B/C ratio of HEAO3~\cite{Engelmann:1990zz} and total $e^\pm$ flux of FERMI~\cite{Ackermann:2010ij}) is given by the dashed brown line.
}
\label{fig:PamelaAMS}
\end{figure}

\begin{figure}[!h]\begin{center}
\includegraphics[width=0.5\textwidth]{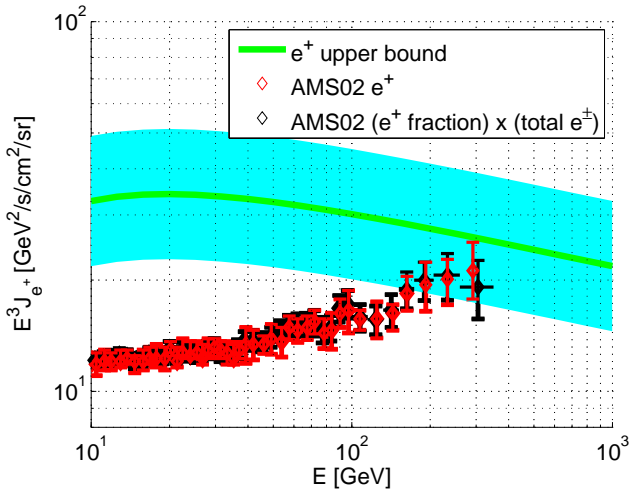}
\end{center}
\caption{Green, with cyan uncertainty band: same as in Fig.~\ref{fig:PamelaAMS}, but showing the $e^+$ flux, rather than the $e^+/e^\pm$ fraction. Red data show the direct AMS02 $e^+$ flux measurement~\cite{AMS02ICRC}. Black data show the $e^+$ flux obtained by multiplying the $e^+/e^\pm$ fraction by the total $e^\pm$ flux, both taken again from AMS02.
}
\label{fig:PamelaAMS0}
\end{figure}

In this paper we point out that the AMS02 measurement~\cite{PhysRevLett.110.141102} is in fact consistent with the simplest possible estimate due to the one guaranteed $e^+$ source: the secondary production of $e^+$ by the collision of high energy primary CRs with ambient interstellar matter (ISM). The main result of this paper is contained in Figs.~\ref{fig:PamelaAMS} and~\ref{fig:PamelaAMS0}. There, AMS02 $e^+/e^\pm$ and $e^+$ data at high energy are seen to comply with an upper bound for secondary production, previously derived in \cite{Katz:2009yd} by ignoring the radiative losses of the positrons. 

In the rest of this paper we outline the derivation of Figs.~\ref{fig:PamelaAMS} and~\ref{fig:PamelaAMS0}, explaining why the AMS02 result provides a strong hint for a secondary positron source.
We comment on the implications of a rising positron fraction, that is not in conflict with a secondary source. Assuming secondary production, we then highlight the constraints imposed by the new measurement on models of CR propagation in the Galaxy.

\noindent
\mysection{AMS02 and the secondary positron flux}
\noindent
%
While the propagation of CRs in the galaxy is poorly understood, the expected fluxes of secondaries, such as positrons, are tightly constrained by the measurement of other secondaries, such as boron. This results from the fact that (i) different relativistic particles with the same rigidity propagate in a magnetic field in the same way, regardless of the magnetic field configuration; and (ii) the production rates of all secondaries are correlated in a calculable manner.

The measured number densities $n_i$ of stable secondary CR nuclei are proportional to their net local production rate and are thus well described by
\begin{equation}\label{eq:Grammage2}
n_i=\frac{X_{\rm esc}\,\sum_{j>i}n_j(\sigma_{j\rightarrow i}/m_p)}{1+(\sigma_i/m_p)X_{\rm esc}},
\end{equation}
where $\sigma_{j\rightarrow i}$ is the decayed spallation cross section of the parent nucleus $j$ into the secondary $i$ per ISM nucleon, $\sigma_i$ is the cross section for destruction of $i$ per ISM nucleon, and $m_p$ is the nucleon mass. The grammage $X_{\text{esc}}$, defined by Eq.~\eqref{eq:Grammage2}, parameterizes the column density of target material traversed by the CRs and is the same for all species. 
Earlier analyses~\cite{1988ApJ...324.1106B,Engelmann:1990zz,2001ApJ...547..264J,2003ApJ...599..582W} relying on HEAO3 data~\cite{Engelmann:1990zz} determined the value of  $X_{\text{esc}}$ to be
\be\label{eq:X} X_{\rm esc}=8.7\left(\frac{E/Z}{10~{\rm GeV}}\right)^{-\alpha}\,{\rm g\,cm^{-2}},\ee
with $\alpha=0.5$ and different fits varying by $\sim 30\%$ in the range $10~{\rm GeV}<E/Z\lesssim100~{\rm GeV}$~\cite{1988ApJ...324.1106B,Engelmann:1990zz,2001ApJ...547..264J,2003ApJ...599..582W}. Here we use new AMS02 B/C data~\cite{AMS02ICRC} to extract the value of $X_{\text{esc}}$ up to $E/Z=1$~TeV. We find (see~\cite{supp}) $X_{\rm esc}$ to be given by eq.~(\ref{eq:X}) with $\alpha=0.4$, slightly harder in slope than the value deduced from the earlier data. 

Eq.~\eqref{eq:Grammage2} does not capture the effect of energy loss during propagation. This means that it cannot be directly applied to positrons, that are subject to Synchrotron and inverse-Compton (IC) losses. Nevertheless, it was realized in~\cite{Katz:2009yd} that Eq.~\eqref{eq:Grammage2} provides a robust upper limit to the positron flux, given that radiative losses can only decrease the flux of the steep positron spectrum. This upper limit is model independent, derived from data and requires no free parameters.

The positron fraction measurements of AMS02~\cite{PhysRevLett.110.141102} and PAMELA~\cite{Adriani:2008zr,Adriani:2010ib} are compared to the upper bound of Eq.~\eqref{eq:Grammage2}, divided by the total $e^\pm$ flux measured by AMS02~\cite{AMS02ICRC}, in Fig.~\ref{fig:PamelaAMS}. 
As mentioned above, AMS02 did not only extend the $e^+/e^\pm$ data to higher energy, it also reported the B/C ratio as well as proton, helium and individual $e^+$ and $e^-$ spectra up to hundreds of GeV to TeV. This enables an improved, compared to what was previously possible, calculation of the $e^+$ upper bound, see Fig.~\ref{fig:PamelaAMS}.
The reported $e^+$ flux~\cite{AMS02ICRC} allows us to compare in Fig.~\ref{fig:PamelaAMS0} the upper bound directly with the data, without involving the $e^-$ flux, which is likely mostly primary and for which there is no definite prediction.

As shown in Figs.~\ref{fig:PamelaAMS} and~\ref{fig:PamelaAMS0}, the upper limit is not violated by the new AMS02 data. This means that the data is consistent with secondary $e+$. Moreover, at high energy the measured $e+$ flux saturates within the secondary limit, previously predicted in~\cite{Katz:2009yd}. This coincidence, while yet to be further tested by future AMS02 data at higher energy, is a compelling hint for a secondary source.

It is worthwhile to compare this result to models invoking new primary sources such as pulsars or dark matter. In such models, ad-hoc tuning of free parameters is required to account for the positron fraction saturating at $\sim0.15$ for $E\gsim200$~GeV. The distinction between the secondary and primary models is even more transparent when considering the absolute $e^+$ flux. In contrast to the $e^+/e^\pm$ fraction, that has a limited dynamical range, the $e^+$ flux due to primary sources could well have been orders of magnitude below or above the secondary bound. We know of no intrinsic scale, and thus of no reason, in any of the primary injection models suggested in the literature, for the $e+$ flux to lie close to the data-driven secondary bound, throughout the range $E\sim10-300$~GeV.

\begin{figure}[!h]\begin{center}
\includegraphics[width=0.5\textwidth]{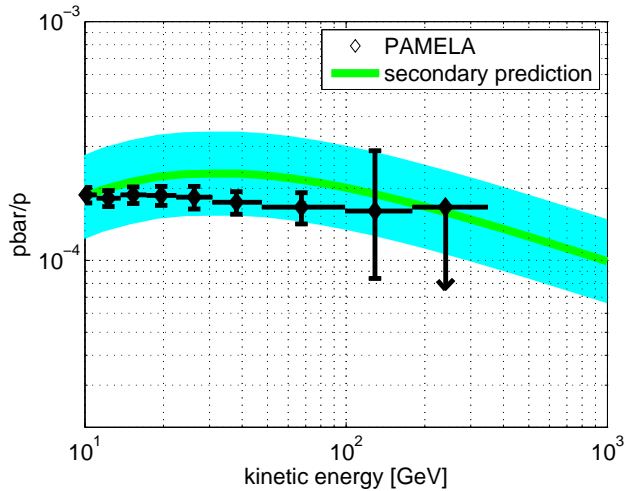}
\end{center}
\caption{PAMELA $\bar p/p$ data~\cite{Adriani:2012paa} vs. the secondary source prediction of Eq.~\eqref{eq:Grammage2}. Cyan shows an estimated calculation uncertainty on the secondary prediction.}
\label{fig:pbar2p}
\end{figure}

\noindent
\mysection{A $\bar{p}$ consistency check, future tests of the secondary model, and calculation uncertainties}\label{intro}
\noindent
A test of the validity of our calculations is presented in Fig.~\ref{fig:pbar2p}, where the measured flux of secondary antiprotons~\cite{Adriani:2012paa}, that are produced in the same interactions as secondary positrons, is compared to the flux obtained from Eq.~\eqref{eq:Grammage2}. As seen in the figure, our calculation is consistent with the observations.

The secondary source hypothesis will be further tested with upcoming AMS02 measurements of the $e^+$ and $\bar p$ flux at higher energy, up to the TeV range~\cite{doi:10.1142/S0218301312300056}. A potentially useful independent check, though complicated by systematic uncertainties, can be done by analyzing the elemental ratios of nuclei \text{having a} radioactive isotope component with a rest frame lifetime of the order of $1~{\rm Myr}$, including Be/B, Cl/Ar and Al/Mg at high rigidity similar to the cooling time of the positrons~\cite{1998ApJ...506..335W,Katz:2009yd,Blum:2010nx}. A more straightforward check, limited however to $E/Z\lsim10$~GeV, will come from directly measuring the isotopic ratio $^{10}$Be/$^9$Be. We note in this context that the early low energy radioactive isotope measurements discussed, for example, in~\cite{2001ApJ...563..768Y}, are limited to $E/Z\lesssim1$~GeV and so cannot be applied model-independently to our study.

We now comment on the systematic uncertainties involved in computing the $e^+$ upper bound and the $\bar p$ flux. 
We estimate these systematic uncertainties roughly by 50\% for both the $e^+$ and $\bar p$ calculations, and denote them by the cyan bands in Figs.~\ref{fig:PamelaAMS}-\ref{fig:pbar2p}. The main potential sources of error are these:

(i) \emph{Different cross section parameterizations} for hadron production in $pp$ and $pA$ collisions, differ by energy-dependent factors in the order of tens of percent. The difficulty is the inapplicability of perturbative calculations, together with the scarcity of accelerator data for soft charged hadron production at high rapidity.
Resolving this ambiguity is beyond the scope of the current paper. Here we follow the same calculation done in~\cite{Katz:2009yd}, to which we refer the reader for more details. 

(ii) \emph{We expect Eq.~\eqref{eq:Grammage2}} to only apply to $\sim10\%$ accuracy, which is roughly the level at which the assumption of negligible energy change during propagation can be expected to hold for stable secondary nuclei. We also estimate about 30\% uncertainty for $X_{\rm esc}$ at 100-500~GeV/nuc.  
Future AMS02 data is expected to significantly improve the determination of $X_{\rm esc}$~\cite{doi:10.1142/S0218301312300056}. While our current parametrization of $X_{\rm esc}$ is consistent with other results~\cite{2008APh....30..133A}, the case is not settled with hints of spectral hardening reported in~\cite{2009AstL...35..338Z,Obermeier:2011wm}.

(iii) \emph{The primary CR nuclei flux and composition} at the 0.1-10~TeV/nuc range, responsible for $\sim10-100$~GeV $e^+$ and $\bar p$ production, are still somewhat uncertain~\cite{Lavalle:2010sf}. Existing measurements at the relevant range~\cite{2011Sci...332...69A,Panov:2011ak,Ahn:2010gv} differ systematically by 20-30\%. In our analysis we adopt a proton flux interpolating the preliminary AMS02 data~\cite{AMS02ICRC}. This data supersedes the earlier PAMELA~\cite{2011Sci...332...69A} and CREAM data~\cite{Ahn:2010gv}, but as for the B/C, we expect significant updates in the near future.

\noindent
\mysection{On a positron fraction rising with energy}
\noindent
Due to synchrotron and IC energy losses, the positron flux is suppressed, compared to the upper bound, by an energy dependent factor $f_{e^+}<1$. $f_{e^+}$ should increase monotonically as a function of $t_c/t_{\rm esc}$, where $t_c$ is the $e^+$ radiative cooling time and $t_{\rm esc}$ is the mean propagation time. In the limit $t_c/t_{\rm esc}\gg1$, we expect $f_{e^+}\to1$.

The claims in the literature, that the increase with energy of the positron fraction is inconsistent with a secondary origin, are based on two lines of reasoning, neither of which is supported by data (see~\cite{Katz:2009yd} for a detailed discussion). The first line of reasoning assumes that (i) primary $p$ and $e^-$ have the same production spectrum, and (ii) primary $e^-$ and secondary $e^+$ suffer the same energy losses. Both (i) and (ii) are unsubstantiated. It is plausible that primary $e^-$ suffer additional energy loss at the primary CR sources and that the injected $e^-$ spectrum is different than that of the protons. The second line of reasoning adopts some specific propagation model, which leads to $t_c/t_{\rm esc}$ (and so to $f_{e^+}$) that decreases with energy. Such behavior of $t_c/t_{\rm esc}$ and $f_{e^+}$ can be modified in alternative models.

This discussion makes clear that given the current AMS02 data, depicted in Figs.~\ref{fig:PamelaAMS}-\ref{fig:PamelaAMS0}, the call for primary sources is unconvincing. Since both $t_c$ and $t_{\rm esc}$ are neither directly measured nor reliably calculable, the energy dependence of $f_{e^+}$, and hence of the positron flux, can not be reliably predicted. The positron data should be regarded as a first direct measurement of $f_{e^+}$, with interesting implications for the times scales $t_{\rm esc}$ and $t_c$ (see e.g.~\cite{Ginzburg:1976dj,Ginzburg:1990sk}, and more recently~\cite{Katz:2009yd,Blum:2010nx}).

\noindent
\mysection{Interpretation: constraints on CR propagation}
In the rest of this paper we assume that the positron flux is of secondary origin, and proceed to deduce new constraints on CR propagation.

The secondary model allows us to quantify the amount by which the positron flux is suppressed by propagation energy loss, based on the observations. The suppression factor $f_{e^+}$ is given by the ratio between the observed $e^+$ flux to the calculated upper bound. This corresponds, in Figs.~\ref{fig:PamelaAMS}-\ref{fig:PamelaAMS0}, to the ratio of the black data to the green curve. We now analyze the constraints arising from Figs.~\ref{fig:PamelaAMS}-\ref{fig:PamelaAMS0}.

\paragraph{1. CR propagation time.} If we ignore Klein-Nishina corrections (see discussion below), then Figs.~\ref{fig:PamelaAMS}-\ref{fig:PamelaAMS0}
imply that:
\be\label{eq:tetc} t_{\rm esc}\left(E/Z=300~{\rm GeV}\right)&\le &t_c\left(E=300~{\rm GeV}\right)\no\\
\label{eq:tetc200}&\sim&1{\rm~Myr}\left(\frac{\bar U_T}{{\rm 1~eVcm^{-3}}}\right)^{-1},\\
t_{\rm esc}\left(E/Z=10~{\rm GeV}\right)&>&t_c\left(E=10~{\rm GeV}\right)\no\\
\label{eq:tetc10}&\sim&30{\rm~Myr}\left(\frac{\bar U_T}{{\rm 1~eVcm^{-3}}}\right)^{-1}.\ee
The RHS of Eqs.~(\ref{eq:tetc200}-\ref{eq:tetc10}) is based on a rough estimate of the $e^\pm$ cooling time at the relevant energies, and as such is subject to O(1) uncertainty.
Here $\bar U_T$ is the time-averaged total electromagnetic energy density in the propagation region. Note that it is natural to expect that $\bar U_T$ should depend on CR rigidity. Thus $\bar U_T$ should be understood as function of $E$, though we omit the explicit dependence for clarity of notation.

One irreducible source for energy dependence in the effective value of $\bar U_T$ comes from Klein-Nishina corrections, that are neglected in Eqs.~(\ref{eq:tetc200}) and~(\ref{eq:tetc10}). 
The Thomson limit is not a good approximation for 20-300~GeV positrons if $\bar U_T$ contains a significant UV component
~\cite{Porter:2005qx}. In that case, the effective radiation energy density for an $\sim300$~GeV $e^+$ can be significantly lower than that for an $\sim10$~GeV one (see e.g.~\cite{Schlickeiser:2009qq,Delahaye:2010ji,2012ApJ...751...71B}).

\begin{figure}[!h]\begin{center}
\includegraphics[width=0.45\textwidth]{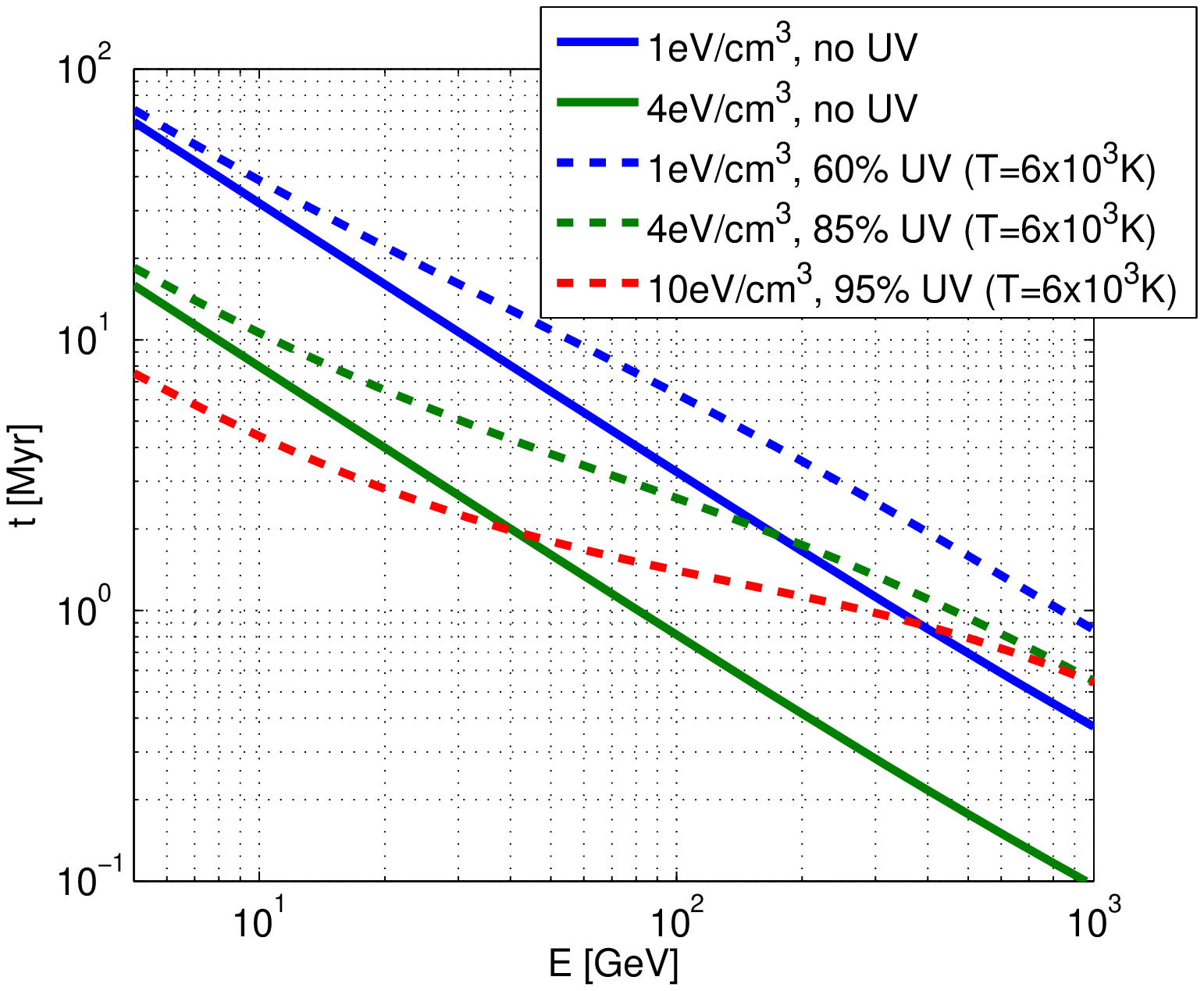}\\
\includegraphics[width=0.45\textwidth]{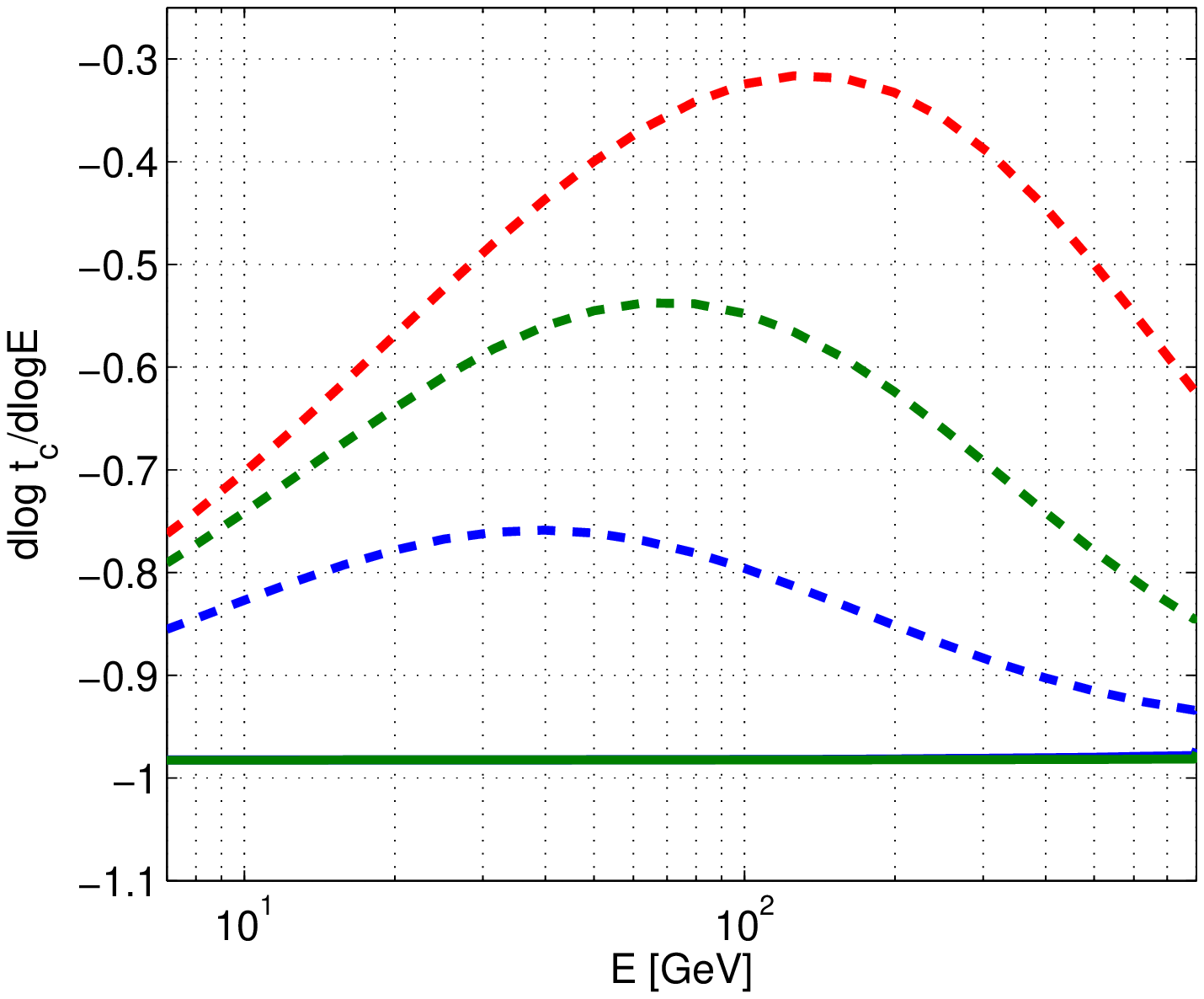}
\end{center}
\caption{Top: cooling time $t_c$ for $e^\pm$ radiative losses, as function of $e^\pm$ energy, for different assumptions regarding the total electromagnetic energy density and its UV component. Bottom: spectral index of the cooling time, color as in top.}
\label{fig:tcool}
\end{figure}

In the top panel of Fig.~\ref{fig:tcool} we plot the cooling time $t_c$ for electrons and positrons under different assumptions for $\bar U_T$. Smooth lines set the UV component to zero. Dashed lines show varying amounts of UV light having a black body spectrum with a temperature $T=6000$~K. The bottom panel shows the spectral index of $t_c$. We learn that significant deviations from the Thomson limit ($d\log t_c/d\log E=-1$) are plausible. In the terms of Eqs.~(\ref{eq:tetc200}-\ref{eq:tetc10}), the effective value of $\bar U_T$ between 10-300~GeV could easily decrease by a factor of 2-3.

We ignore bremsstrahlung (brem) and adiabatic losses. The brem optical depth can be estimated as $\tau_{\rm brem}\sim X_{\rm esc}/\zeta\approx0.1\left(E/20~\rm{GeV}\right)^{-0.4}$, where $\zeta\sim60$~g/cm$^2$ is the electron radiation length, and is too small to explain the $e+$ loss inferred from Fig.~\ref{fig:PamelaAMS0}.  Adiabatic loss applies equally to $e+$ and $\bar p$, and is thus constrained to be small by the $\bar p$ flux.

\paragraph{2. The mean ISM density of the CR halo.} We can now estimate the mean ISM density traversed by CRs. Using Eq.~(\ref{eq:X}) together with Eqs.~(\ref{eq:tetc200}) and~(\ref{eq:tetc10}), we find
\be\bar n_{ISM}\left(E/Z=300\,{\rm GeV}\right)\label{eq:n200}&\gtrsim&1\,\left(\frac{\bar U_T}{{\rm 1~eVcm^{-3}}}\right)\,{\rm cm^{-3}},\;\;\;\\
\bar n_{ISM}\left(E/Z=10\,{\rm GeV}\right)\label{eq:n10}&\lesssim&0.15\,\left(\frac{\bar U_T}{{\rm 1~eVcm^{-3}}}\right)\,{\rm cm^{-3}},\;\;\;
\ee
assuming ISM composition of 90\%H+10\%He by number.

Eqs.~(\ref{eq:n200}) and~(\ref{eq:n10}) suggest that the confinement volume of CRs decreases with increasing CR rigidity, to the extent that CRs at $E/Z\sim300$~GeV spend much of their propagation time within the thin Galactic HI disc, with a scale height $h\simeq200\,{\rm pc}$, while CRs at $E/Z\sim10$~GeV probe a larger halo. These are not robust conclusions, however. For example, if a significant fraction of the grammage $X_{\rm esc}$ is accumulated during a short time in dense regions, e.g. near the CR source~\cite{Cowsik:2010zz}, then the halo could be larger. Energy dependence in $\bar U_T$ could further affect the interpretation. For example, $\bar U_T\propto E^{-0.6}$ (inspired by the CR grammage $X_{\rm esc}\propto E^{-0.4}$) would allow for a rigidity independent $\bar n_{ISM}$.

Finally, we comment that a rising $f_{e^+}$ is comfortably compatible with the observed primary proton spectrum $J_p\propto E^{-2.8}$. It is clear from Fig.~\ref{fig:tcool}, that $t_{\rm esc}$ falling as $E^{-0.8}$ or so could lead to $t_c/t_{\rm esc}$, and thus to $f_{e^+}$, that rise with increasing energy. Consider first the possibility that the CR halo decreases with increasing energy. As an example along this line~\cite{Katz:2009yd}, one-dimensional diffusion, with null boundary conditions at a CR scale height $L\propto E^{-0.4}$, and rigidity-independent diffusion coefficient, would give $t_{\rm esc}\propto E^{-0.8}$, flat or rising $f_{e^+}$, and $X_{\rm esc}\propto E^{-0.4}$, consistent with observations. In this case the inferred proton injection spectrum would be $\propto E^{-2.4}$. If, on the other hand, CR confinement occurs at fixed volume, then the proton index could be interpreted as $E^{-0.8}$ softening by escape, on top of an $E^{-2}$ injection. In this case, the slope of $X_{\rm esc}\propto E^{-0.4}$ would imply that the CR distribution is not homogeneous in the spallation region, with possible ramifications for gamma ray observations.

\noindent
\mysection{Conclusions}
The positron fraction measured by AMS02 is consistent with the upper bound predicted in Ref.~\cite{Katz:2009yd}, assuming a secondary source. Upcoming AMS02 measurements of the $e^+$ and $\bar p$ flux at yet higher energies will continue to test the model.

At the highest measurement energy, the positron flux saturates the upper bound, and throughout the measurement range it is never smaller than a factor of $f_{e^+}\sim0.3$ compared to it. We find this to be a compelling hint for a secondary source. Considering hypothetical primary sources such as pulsars or dark matter, we know of no intrinsic scale in these models that would fix the positron flux at this particular range.

Interpreted under the secondary source hypothesis, the positron data places interesting constraints on the propagation time of CRs at $E/Z\sim10-300$~GeV, that we roughly summarize by $t_{\rm esc}(10~{\rm GeV})\gtrsim30$~Myr and $t_{\rm esc}(300~{\rm GeV})\lesssim1$~Myr. The constraint on $t_{\rm esc}$ at $E/Z>100$~GeV is obtained by the new positron data, with no direct counterpart in earlier CR  data. The constraint at $E/Z\sim10$~GeV is consistent, within uncertainties, with measurements of the elemental ratios of radioisotopes~\cite{1998ApJ...506..335W,Katz:2009yd,Blum:2010nx}.

Using the measured CR grammage together with the new constraints on $t_{\rm esc}$, we derive the mean ISM particle density in the propagation region of high energy CRs, $\bar n_{ISM}\gtrsim1$~cm$^{-3}$ for $E/Z=300$~GeV. This result for $\bar n_{ISM}$ is comparable to the mean ISM density in the Milky Way HI disc.
At $E/Z=10$~GeV we find a smaller mean density, $\bar n_{ISM}\lesssim0.15$~cm$^{-3}$. Put together, these numbers could mean that the scale height of the CR halo decreases with increasing CR rigidity
(however, see discussion following Eqs.~\ref{eq:n200} and~\ref{eq:n10} for alternative interpretations).

\mysections{Acknowledgments}
We thank Moti Milgrom, Kohta Murase and Rashid Sunyaev for discussions.
KB is supported by DOE grant DE-FG02-90ER40542. BK is supported by NASA through the Einstein Postdoctoral Fellowship awarded by Chandra X-ray Center, which is operated by the Smithsonian Astrophysical Observatory for NASA under contract NAS8-03060. EW is partially supported by GIF and UPBC grants.

\mysections{Supplementary material: CR grammage from AMS02 data}
AMS02 provided a new measurement of the B/C ratio up to rigidity of approximately one TV. This measurement, combined with high energy primary-to-primary flux ratios from other experiments, can be used to infer the CR grammage up to the same rigidity, allowing us for the first time to predict the secondary $e^+$ flux bound and $\bar p$ flux all the way to the TeV.

\begin{figure}[!h]\begin{center}
\includegraphics[width=0.5\textwidth]{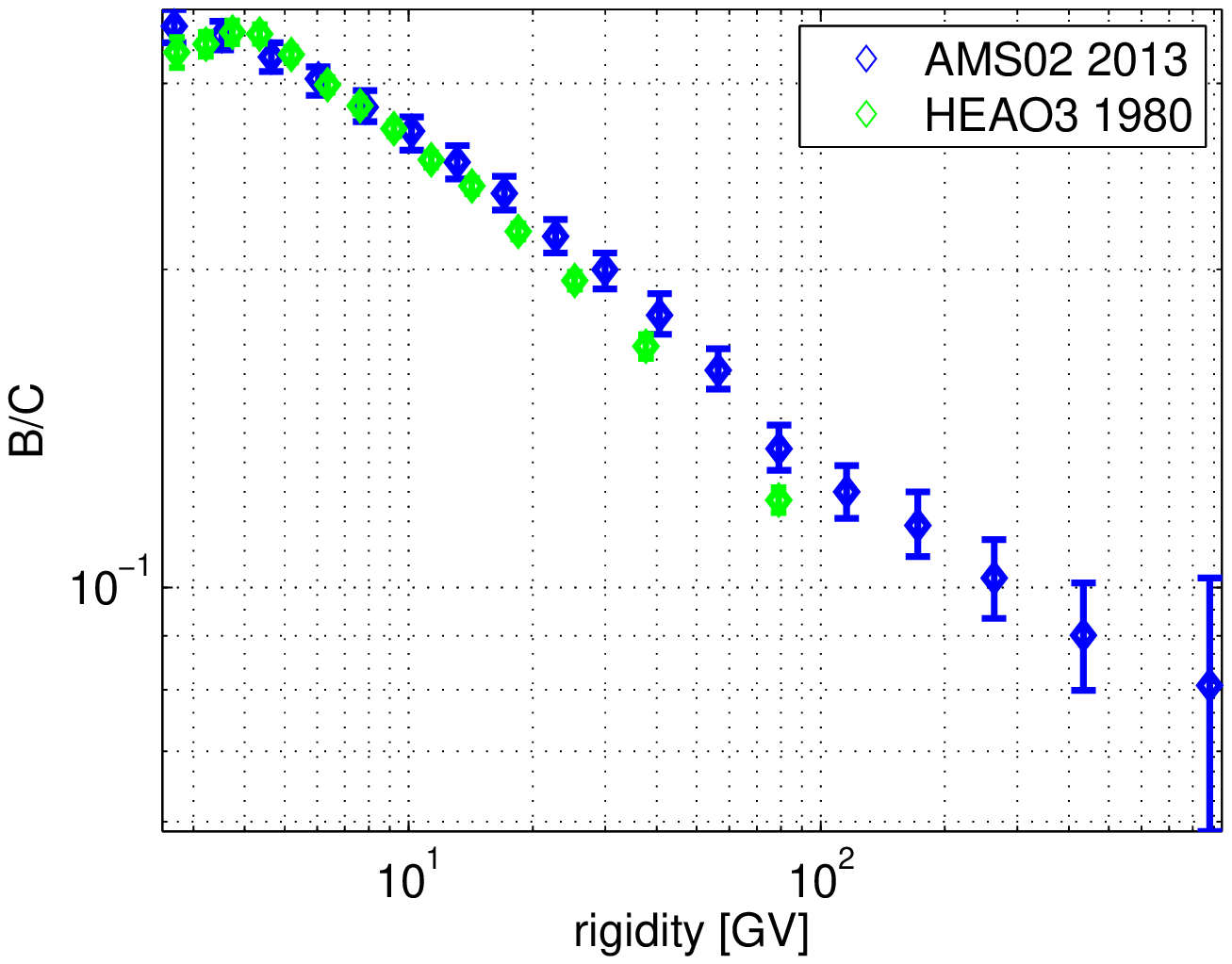}\\
\includegraphics[width=0.5\textwidth]{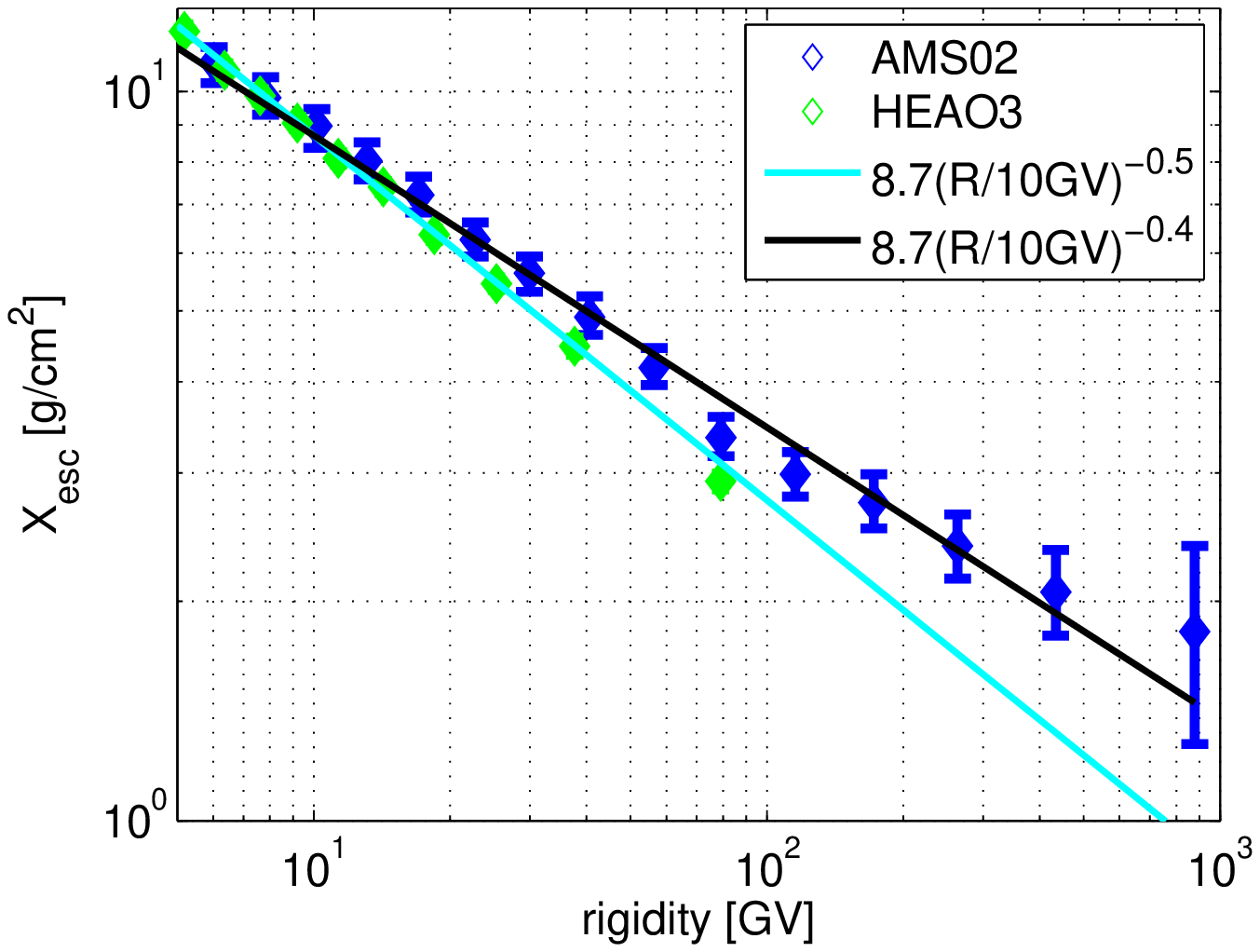}
\end{center}
\caption{Top panel: B/C measured by HEAO3~\cite{Engelmann:1990zz}(green) and AMS02~\cite{AMS02ICRC}(blue) as function of rigidity. Bottom panel: CR grammage deduced from Eq.~(\ref{eq:XBC}), with fitting formulae.
}
\label{fig:B2C}
\end{figure}

In the top panel of Fig.~\ref{fig:B2C} we show the new AMS02 B/C measurement, together with the earlier HEAO3 data, as function of rigidity. Measurements of primary-to-primary flux ratios, including C/O, N/O, and Fe/O up to the TV range were reported by the CREAM experiment~\cite{2010ApJ...715.1400A}. These data are in good agreement with straight-forward extrapolation of the earlier lower energy HEAO3 data~\cite{Engelmann:1990zz}, and suggest that for $E/Z>10$~GeV the quantity
\be\sum_{_{i>B}}\left(n_i/n_C\right)\sigma_{i\to B},\ee
representing the production term for boron divided by the carbon flux, is independent of rigidity to a very good approximation. This fact, together with the agreement between the HEAO3 and AMS02 B/C data at $E/Z\sim10$~GeV, allows us to use the $X_{\rm esc}$ parametrization of~\cite{2003ApJ...599..582W}  in the vicinity of $E/Z=10$~GeV in order to solve for $X_{\rm esc}$ as function of B/C at higher rigidity:
\be\label{eq:XBC}X_{\rm esc}\approx\frac{20\left(n_B/n_C\right)}{1-1.5\left(n_B/n_C\right)}\,{\rm g\,cm^{-2}}.\ee

In the lower panel of Fig.~\ref{fig:B2C} we show the resulting $X_{\rm esc}$, obtained by using Eq.~(\ref{eq:XBC}) with HEAO3 and AMS02 data. We find that AMS02 data is well fit by a power law, given by 
\be\label{eq:X} X_{\rm esc}=8.7\left(\frac{E/Z}{10~{\rm GeV}}\right)^{-\alpha}\,{\rm g\,cm^{-2}},\ee
with $\alpha=0.4$.

\bibliography{ref_v2}

\end{document}